\documentclass[12pt,preprint]{aastex}

\slugcomment{Submitted to AJ (revised version)}

\shorttitle{}
\shortauthors{Martin}

\begin{document}


\title{A New Multiple Stellar System in the Solar Neighborhood.}


\author{Eduardo L. Mart\'\i n}
\affil{Institute for Astronomy, University of Hawaii at Manoa, 
2680 Woodlawn Drive, Honolulu HI 96822, USA}
\email{ege@ifa.hawaii.edu}



\newpage

\begin{abstract}
 
Adaptive optics corrected images obtained with the 
CIAO instrument at the Subaru 8.2-meter 
telescope show the presence of two subarsecond companions to 
the nearby (d=19.3~pc) young star GJ~900, 
which was previously classified as a  
single member of the IC~2391 supercluster. 
The two companions share the same proper motion as the 
primary and are redder. Their projected separations from the 
primary are 10~AU and 14.5~AU for B and C, respectively. 
The estimated masses for the two new companions 
depend strongly on the age of the system. 
For the range of ages found in the literature for IC~2391 
supercluster members (from 35~Myr to 200~Myr), the expected masses 
range from 0.2~M$_\odot$ to 0.4~M$_\odot$ for the B component, and from 
0.09~M$_\odot$ to 0.22~M$_\odot$ for the C component.  
The determination of the dynamical mass  
of the faintest component of GJ~900 will yield  
the age of the system using theoretical evolutionary tracks. 
The apparent separations of the GJ~900 system 
components meet the observational criterion for an    
unstable Trapezium-type system, but this could be a projection effect. 
Further observations are needed to establish the nature of this 
interesting low-mass multiple system.  

 
\end{abstract}




\section{Introduction}

The star GJ~900 (BD+00~5017, HIP~116384) 
is one of the nearest (d=19.3$\pm$0.6~pc; Perryman et al. 1997) cool 
(K7-M0) young stars known. It has been remarked because of its high level 
of H$_\alpha$ and CaII H+K emission (Stauffer \& Hartmann 1986; Bopp 1987;  
Giampapa, Cram \& Wild 1989), as well as strong  
X-ray emission (Micela, Favata \& Sciortino 1997; 
Sterzik \& Schmitt 1997), which indicate  
a young age. Montes et al. (2001) classify it as a member of the 
IC~2391 supercluster kinematic group. The IC~2391 cluster has 
a classical (isochrone fitting) age of 35~Ma\footnote{1 Ma = 10$^6$ years}, 
which has been recently revised 
to 53~Ma using lithium observations of very low-mass (VLM) cluster members 
(Barrado y Navascu\'es, Stauffer, \& Patten 1999). 
On the other hand, Eggen (1991) found that IC~2391 supercluster stars have a bimodal 
age distribution (80~Ma and 250~Ma). 

The far-infrared fluxes of GJ~900 are normal for its spectral type 
(Mathioudakis \& Doyle 1993), so there is no evidence for a large amount 
of warm dust in the system. Bopp (1987) reported 15 radial velocity measurements 
for GJ~900 obtained over a 3 year baseline. 
There was no sign for radial velocity variability larger than the 
error bars (1~$\sigma$=1~km s$^{-1}$). Bopp also reported that GJ~900 
V-band photometry was constant within 0.01 magnitudes at V=9.54. 
Additionally, Bessell (1990) found V=9.57. Hence, there is no evidence 
for large photometric variability despite of its high level of 
chromospheric and coronal activity. 

We included  GJ~900 in an ongoing high-resolution imaging survey for brown dwarf 
companions to nearby young cool stars. The star was thought to be single, 
but our observations have revealed the presence of two VLM 
companions. The projected separations are such that GJ~900 could be a rare 
example of a young unstable Trapezium triple system. Monitoring of the orbital 
motion of this system is needed to determine the true separations of 
the components. In Section~2, we present our observations and data analysis. 
In Section~3, we discuss the results and estimate masses for the components 
using theoretical models. 
 
\section{Observations and Data Analysis}

GJ~900 was observed in two observing runs (7 August 2002 and 18 January 2003) 
with the Coronagraphic Imager with Adaptive Optics (CIAO) 
attached to the 8.2-meter Subaru telescope (see Tamura et al. 2000 for a 
description of the instrument). 
We used an AO imaging mode with a pixel scale of 0$\arcsec$.022 pix$^{-1}$ 
and a field of view of 22.3 square arseconds. 
We did not use the coronograph. It was inmediately apparent 
in the first inspection of the AO-corrected 
images at the telescope that there were two companions 
to GJ~900. 

In the first run, we took 6 images of 50~s, 6 images of 20~s, 18 
images of 3.3~s in the H-band, and 6 images of 3.6~s in the K-band. 
In the second run, we took  
12 images of 3~s in the H-band. The natural seeing in the second run  
was much worse than in the first one. 
Data reduction was performed using standard 
IRAF\footnote{IRAF is distributed by National Optical Astronomy Observatories, 
which is operated by the Association of Universities for Research in Astronomy, Inc., 
under contract with the National Science Fundation.}
 routines which 
included bias subtraction, cosmic ray correction, and flat fielding. 
Differential magnitudes were calculated 
using the \emph{phot},  \emph{psf} and \emph{allstar} routines in 
IRAF's \emph{daophot} package. 
The results are given in Table~1, together with relative astrometry of the 
triple system's components. Error bars were estimated from the 
dispersion of the measurements obtained from individual images.

\section{Discussion} 

The baseline of our two epochs is 163 days. The proper motion of 
GJ~900 is 344 mas/yr, so the star should have moved 155 mas (7 pixels) 
with respect to any background star. We would have detected 
such motion very easily in our images. Thus, we confirm that 
the two companions share the same proper motion as the primary. 
Moreover, there were not any other stars in the field of view of 
the CIAO detector, and the two companions are redder in H-K 
than the primary (Table~1). 

The near-infrared magnitudes of GJ~900 (combined flux of the whole system) 
are J=6.865$\pm$0.021, H=6.245$\pm$0.007 and K=6.008$\pm$0.014 
in the TCS system (Alonso, Arribas \& Mart\'\i nez-Roger 1994). 
Using these photometric data and the differential photometry obtained 
from our AO observations, we derive H=6.52 for GJ~900~A, 
H=8.30 for GJ~900~B, and H=9.07 for 
GJ~900~C. The distance modulus is m-M=1.46$\pm$0.03, and thus the 
absolute magnitudes are H=5.06, 6.84 and 7.61 for GJ~900~A, B and C, 
respectively. 
In order to estimate the masses of the components, we need to know 
the age of the system.  Montes et al. (2001) have argued that it 
belongs to the IC~2391 supercluster on the basis of its space motion. 
IC~2391 cluster's classical age is $\sim$35~Ma. 
However, the age of IC~2391 has recently been revised to 53~Ma 
using the lithium depletion boundary method (Barrado y Navascu\'es 
et al. 1999).   

For an age of 35~Ma, we derive the following masses from 
the absolute H-band magnitudes and models tailored for GJ~900 provided 
by I. Baraffe (they use the input physics described in 
Baraffe et al. 1998): 
0.60~M$_\odot$, 0.20~M$_\odot$ and 0.09~M$_\odot$ for GJ~900~A, B and C, 
respectively. For an age of 50~Ma, we infer masses of 
0.61~M$_\odot$, 0.22~M$_\odot$ and 0.14~M$_\odot$ 
using the same models. As expected, the mass of the lowest mass 
component is the most sensitive to the age of the system. 
Older ages, such as those proposed by Eggen (1991), lead to higher masses 
for the C component. In the referee report for this paper, 
John Stauffer has argued that GJ~900~A 
has an age of about 100~Ma or older on the basis of its position in the 
$V$ versus $V-I$ color-magnitude diagram after correcting the flux for binarity 
and after comparing with low-mass members of the IC~2391, IC~2602 and Pleiades 
open clusters. The correction for the presence of the companions at visual 
wavelengths remains speculative until optical data is acquired for each of them. 
Such a study is out of the scope of this paper.  
For an age of 100~Ma, the masses of B and C would be 
0.36~M$_\odot$ and 0.21~M$_\odot$, respectively. 

Determination of dynamical masses in the future will yield an improved 
age for the system using evolutionary models. In order to get a rough 
estimate of what to expect for the orbital period, we assume an age of 
50~Myr and a total mass of the system of 0.97~M$_\odot$. For a distance 
of 19.3~pc, 
the projected separation between A and B is 10~AU, and 
between A and C is 14.5~AU. Thus, we estimate orbital periods of 
$\sim$34 years for GJ~900~B and $\sim$56 years for GJ~900~C. 

A hint of orbital motion may already be present in the difference 
between the position angles measured in the two different epochs 
of CIAO observations (Table~1). Component B  has moved 2.9 degrees 
in 0.45 years, while component B has not moved more than 0.2 degrees. 
If the orbits are circular, the orbital periods would be $\sim$56 years 
for B, and $\ge$810 years for C, respectively. 
These periods are longer than our previous estimate, but we should bear 
in mind that most binary orbits are quite eccentric, and the components 
spend a larger fraction of the time near periastron. 
We will have to cover a significant fraction of the orbits to derive 
reliable orbital parameters. 

\subsection{Is GJ~900 a Trapezium-like multiple system?}

Trapezium-like multiple systems have been defined as physical 
systems with a least 3 components where the largest separation is 
less than 3 times the smallest separation (Abt \& Corbally 2000).  
The apparent configuration of the GJ~900 system meets this criterion because 
the largest separation is 1.45 times the smallest one. 

The Abastumi Catalogue of Trapezium-type multiple systems 
includes 637 objects. The masses of all these objects are several 
times that of the Sun, and the ages are typically younger 
than 50~Ma (Abt \& Corbally 2000; Allen \& Poveda 1975; Harrington 1992; 
Salukvadze 1999, 2000) because they are thought to be unstable. 
Because of its present configuration, it appears that GJ~900 
could be the first example of a sub-solar mass 
Trapezium-type multiple system. Its survival would indicate an age 
younger than $\sim$40~Ma, or a peculiar orbital stability 
(resonances). 

If GJ~900 is significantly older than 40~Ma, it is likely to be stable, 
and hence not a true Trapezium system. 
There is about 10\% chance that GJ~900 looks like 
a Trapezium-type system because of a chance projection 
(Ambartsumian 1954). Astrometric monitoring 
of the orbital motion is required to determine the inclination 
of the orbit with respect to the line of sight. In a few years, 
we should be able to tell whether GJ~900 is a Trapezium-like system 
or not, to estimate dynamical masses for the components, and 
to derive an age for the system.

\acknowledgments

I thank the Subaru team for their assistance: Sumiko Harasawa, Koji Murakawa, 
Shin Oya, and Tom Winegar. I also thank Laird Close, Melanie Freed, Tim Kendall 
and Nick Siegler for participating in the CIAO observing runs, Sharon Velez Erickson 
for assisting with data reduction, Isabelle Baraffe for providing models tailored 
for GJ~900, and John Stauffer for a helpful referee report. 
Support for this work was provided by National Aeronautics and Space
Administration(NASA) grant NAG5-9992 and National Science Foundation(NSF)
grant AST-0205862. 
This research has made use of the SIMBAD database, operated at CDS, Strasbourg, France. 
The author wish to extend special thanks to those of Hawaiian ancestry on 
whose sacred 
mountain of Mauna Kea we are privileged to be guests. 
Without their generous hospitality, 
the Subaru telescope observations presented therein would 
not have been possible.

\clearpage


\begin{figure}
\plotone{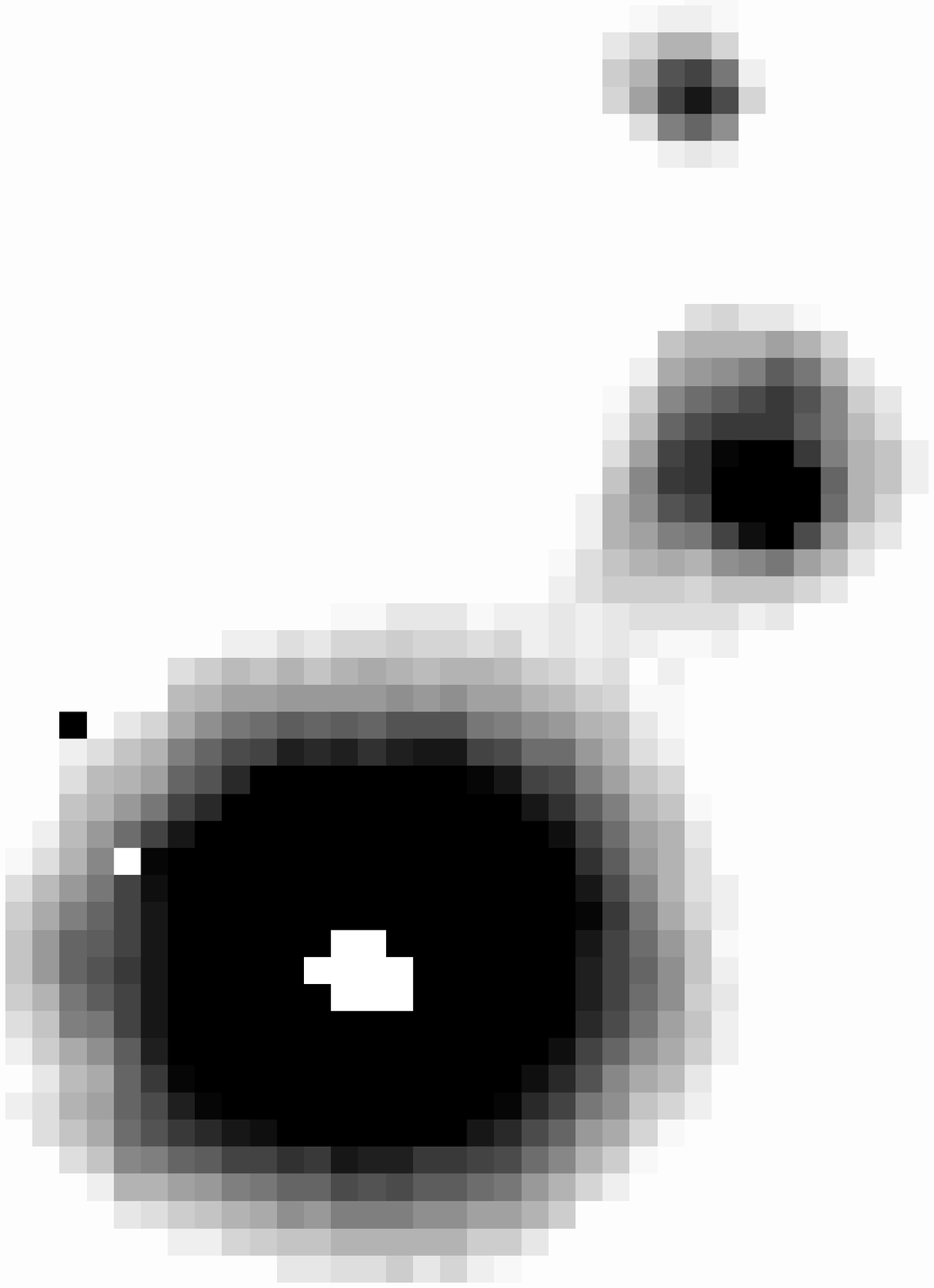}
\caption{CIAO discovery image of the GJ~900 triple system obtained in H-band on 
7 August 2002. North is up and east to the left. The angular separation 
between A and B is 0$\arcsec$.51, and the separation 
between A and C is 0$\arcsec$.76. 
GJ~900 is a low-mass Trapezium-type triple system candidate. }
\end{figure}

\clearpage

\begin{deluxetable}{lllllllllll}
\tabletypesize{\scriptsize}
\tablecaption{CIAO data for GJ~900 A,B,C.}
\tablewidth{0pt}
\tablehead{
\colhead{Epoch} & 
\colhead{Components} &
\colhead{Delta H} & 
\colhead{Delta K} &
\colhead{Sep.} &
\colhead{PA}  
 }
\startdata
7 Aug 2002      & A-B & 1.78$\pm$0.02 & 1.61$\pm$0.03 & 0$\arcsec$.51$\pm$0.01 & 324.5$\pm$0.1  \\ 
7 Aug 2002      & A-C & 2.55$\pm$0.03 & 2.38$\pm$0.04 & 0$\arcsec$.76$\pm$0.01 & 344.0$\pm$0.1   \\ 
18 January 2003 & A-B & 1.70$\pm$0.04 &               & 0$\arcsec$.52$\pm$0.02 & 327.4$\pm$0.1   \\ 
18 January 2003 & A-C & 2.31$\pm$0.06 &               & 0$\arcsec$.74$\pm$0.02 & 343.9$\pm$0.1   \\
\enddata
\end{deluxetable}

\end{document}